# Collective quantum states at the atomic limit

Fan Zhang[1†], Yanxing Li[1†],Chengye Dong[2†], Ninad Kailas Dongre[3], Viet-Anh Ha[1,4], Yu-Chuan Lin[2,5], Yiyuan Luo[1], Hyunsue Kim[1], Joshua A. Robinson[2], Feliciano Giustino[1,4], Fan Zhang[3], and Chih-Kang Shih[1*]

[1]Department of Physics, University of Texas at Austin, Austin, TX, USA.

[2]Department of Materials Science and Engineering, Pennsylvania State University, University Park, PA, USA

[3]Department of Physics, University of Texas at Dallas, Dallas, TX, USA.

[4]Oden Institute for Computational Engineering and Sciences, University of Texas at Austin, Austin, TX, USA.

[5]Department of Electrophysics, National Yang Ming Chiao Tung University (NYCU), Hsinchu, Taiwan

## Abstract

Collective quantum states are often associated with extended systems, where spatially extensive degrees of freedom enable emergent many-body behavior; whether such strongly correlated states survive at atomic dimensions remains a fundamental question. Tomonaga–Luttinger liquids provide a paradigmatic example of one-dimensional collective quantum matter characterized by spin–charge separation. Using low-temperature scanning tunneling microscopy and spectroscopy, we directly visualize quantized collective modes in atomically confined mirror twin boundary segments of monolayer $WSe_2$. Distinct standing-wave branches associated with fractionalized spin and charge excitations persist in segments as short as one nanometer, establishing the atomic-scale confinement limit of Luttinger-liquid behavior. These ultrashort segments form a new class of many-body quantum dots whose discrete spectra arise from confined collective bosonic modes rather than single-particle electron states. When assembled into ordered chains, inter-dot coupling reshapes electron-like fundamental states while collective spin/charge excitations remain largely intact, revealing distinct coupling responses of emergent many-body modes. Our results demonstrate that collective quantum matter can persist and exhibit fundamentally distinct coupling behavior at atomic length scales, establishing a novel platform for engineering strongly correlated quantum phases from atomically confined building blocks.

## Main

Collective quantum behavior in condensed matter is typically associated with extended systems, where long-wavelength correlations and collective excitations emerge from spatially extensive degrees of freedom[1,2]. As system size approaches atomic dimensions, however, discrete quantization, structural localization, and enhanced boundary effects are generally expected to disrupt such collective phenomena[3–10]. Understanding whether defining signatures of correlated quantum matter can persist at these extreme spatial limits represents a fundamental challenge, particularly in one-dimensional systems where interactions dominate the low-energy physics[11,12].

Tomonaga–Luttinger liquid (TLL) theory[13–16] provides a paradigmatic framework for understanding collective behavior in one-dimensional interacting systems, predicting spin–charge separation and non-Fermi-liquid excitations[17–26]. As illustrated schematically in Fig. 1a, an injected electron fractionalizes into separate density waves: a charge excitation (holon) and a spin excitation (spinon) that propagate with different velocities. In an infinitely long TLL (Fig. 1b), these modes travel freely without reflection, producing spatially separated spin and charge density waves whose velocities are set by the interaction strength. In contrast, a finite-length TLL, such as the mirror twin boundary (MTB) [27–40] in monolayer transition metal dichalcogenides (Fig. 1c), the confinement leads to distinctive standing waves for spin and charge modes (Fig. 1d). More strikingly, the confinement introduces an energy gap $\Delta E_g = E_0 + E_C$, where $E_0 = \hbar v_F \pi / L$ ($\hbar$ is the reduced Planck constant and $L$ the MTB length) is the single-particle level spacing, and $E_C$ accounts for the electron–electron repulsion. (Fig. 1d). Such standing waves for spin/charge modes and the correlation gap have been directly imaged using scanning tunneling spectroscopy in finite-length MTBs[28–31] in materials such as $MoSe_2$, $WS_2$, and $MoS_2$. However, these investigations have largely focused on segments with lengths sufficient to sustain well-developed collective behavior (i.e., $L \gg \lambda_F$), leaving open the fundamental questions of how far such physics can persist as quantum confinement approaches atomic dimensions, and how interacting arrays of such confined units behave when coupled together.

Here we address these two fundamental questions by engineering atomically confined MTB segments that reduce the effective length of a one-dimensional system to the nanometer scale. This platform enables a direct test of the boundary between few-body atomic confinement and many-body collective behavior, allowing us to determine whether spin–charge-separated excitations persist as spatial extent approaches atomic dimensions.

### TLL behavior in $WSe_2$ mirror twin boundaries

We first benchmark the behavior of TLL in *conventional MTB* in $WSe_2$ monolayers. Shown in Fig. 2a is the density functional theory (DFT) calculation of single-particle band structure where the MTB states are marked by orange dots. Indeed, near the Fermi energy, the MTB states exhibit a linear dispersion (highlighted by the black dashed line) with a group velocity of $v_F = 2.17$ eV · Å. Shown in Fig. 2b is the STM image of a 5 nm long MTB with spatially resolved STS shown in Fig. 2c. The spectra reveal multiple confined states throughout the measured energy window, separated by a narrow energy gap of $\Delta E_g = 300$ meV. All states reside within the 2.2 eV bandgap of the surrounding semiconducting $WSe_2$, confirming the localization of MTB and its topological origin[33].

To visualize the spatial modulation of these states, we plot a color map of the STS signal as a function of position and sample bias in Fig. 2d. Several key features emerge: (i) the gap between the highest occupied state (HOS) and the lowest unoccupied state (LUS) is spatially uniform; (ii) multiple energy levels appear with varying level spacings; and (iii) the HOS and LUS exhibit out-of-phase spatial profiles, while the first spin (S1) and charge states (C1) display in-phase modulations. Also, from the energy levels of first spin and charge modes, the Luttinger liquid interaction parameter is estimated to be $K_c = 0.46$, indicating a strongly interacting TLL regime (Details of the estimation are described in Supplementary Note 1). These observations are well captured by the "TLL-in-a-box" model[28,31], and further supported by simulated local density of states (LDOS) from a finite-length TLL calculation shown in Fig. 2e. Additional evidence for finite-size TLL behavior is provided in Fig. 2f, where the energy gap $\Delta E_g \propto 1/L$ is confirmed by fitting data from MTBs of varying lengths. Moreover, as illustrated in bias-dependent dI/dV mappings (Supplementary Fig. S1), each successive energy level is associated with one additional node, consistent with the predicted quantization in a 1D box.

As confinement approaches atomic length scales, interpreting discrete excitation spectra becomes increasingly challenging because single-particle quantization and structural localization can produce features superficially similar to collective modes. Furthermore, conventional expectations based on spatial extensivity formulations of Tomonaga–Luttinger liquid theory suggest that fractionalized spin and charge excitations should cease to be well defined when the system length approaches atomic scales[3,6–10]. Establishing the persistence of collective behavior in this regime therefore requires identifying signatures that cannot be explained by single-particle confinement alone. In the following, we demonstrate that distinct excitation branches with characteristic phase relationships and interaction-dependent energy scaling remain intact even at nanometer confinement, providing strong evidence for the survival of collective quantum modes.

**Collective quantum behavior in atomic-limit boundaries**

Having established finite-length TLL behavior, we next investigate how far collective excitations can persist as the system size approaches atomic dimensions. By fabricating MTB segments with lengths approaching one nanometer, we probe the ultimate spatial limit of spin–charge separation.

In Fig. 3a, the repetitive arrangement of the short segment MTB and the 8-fold ring leads to a $\sqrt{21}a$ commensurability ($4.5a$ along zigzag and $\sqrt{3}/2a$ along armchair). Also shown are STM images acquired at a sample bias of -0.4 V (Fig. 3b) and 0.8 V (Fig. 3c) respectively. In Fig. 3b, the bias is in the band gap and tunneling occurs from the underlying graphene, leading to prominent moiré pattern due to the 3:4 coincident lattice match between $WSe_2$ and graphene. At the anti-phase boundary (APB) (see more STM images of typical APBs in Supplementary Figures S2 and S3), the short MTB appears as a straight line while the kink demarks the 8-fold ring. Note that at every other kink, the geometric phase matches with the underlying graphene, resulting in an elongated bright feature (further discussed in Supplementary Figure S4). Outside the energy gap $\Delta E_g$, the TLL state dominates, and the STM image shows spatially extended features spanning across a region of ~ 1 nm perpendicular to the MTB.

The electronic properties are revealed by spatially resolved dI/dV spectra along (Fig. 3e) and perpendicular (Fig. 3f) to the chain (Fig. 3d). Remarkably, the superlattice retains key signatures of TLL physics. First, a large apparent gap of $\Delta E_g$= 1.18 eV is observed (marked as a red star in Fig. 2f). This value is slightly smaller than a value of 1.4 eV based on Fig. 2f, extrapolated to L = 1 nm (the length of MTB). This is attributed to the penetration of TLL states into the 8-fold ring that isolates them. Most remarkably, the spin–charge separation is preserved: the zero mode (LUS) and first excitation mode (S1) appear out of phase, while the first spin (S1) and charge (C1) excitations are in phase. These observations confirm that the system behaves as a coupled array of TLL quantum dots. From the excitation energy levels, one can deduce an effective Luttinger parameter of $K_c = \frac{v_s}{v_c} = 0.55$, slightly higher than the value of $K_c = 0.46$ for longer MTBs, indicating an effectively weaker intra-dot interaction as a result of the interplay between atomic confinement and inter-dot couplings/interactions in this linear chain.

**Coupled array of many-body quantum dots at the atomic limit**

Having demonstrated that collective modes survive at the atomic confinement limit (Fig. 3), we now examine how such confined segments behave when coupled into an ordered chain. This configuration provides a direct test of whether electron-like and collective excitations respond differently to inter-dot coupling[41–48].

To establish the single-particle expectation for the superlattice geometry, we first compare energy-resolved tunneling spectra with DFT calculations (Figs. 4a, b). The DFT results provide a baseline for the non-interacting electronic structure, identifying a localized in-gap state associated with the eight-membered ring (peak 1) as well as several states confined within the short MTB segments (peaks 2–4). However, the STM/S measurements reflect the interacting spectral function; consequently, spatial redistribution and hybridization observed experimentally need not follow the single-particle density of states predicted by DFT. Experimentally, a pronounced spectral feature near −0.7 V (black arrow) closely matches the calculated ring-localized state (Peak 1), confirming its structural origin and serving as a single-particle reference for comparison. In contrast, the remaining experimentally observed excitations exhibit substantial deviations from the single-particle predictions. In fact, these deviations are consistent with interaction-driven renormalization of the spectral function, as expected in one-dimensional correlated systems.

Spatially resolved spectroscopic imaging (Fig. 4c–4l) reveals a clear distinction between the behavior of fundamental electron-like states and collective excitations within the coupled chain. The −0.7 V state is strongly confined to the eight-membered ring, consistent with its single-particle character predicted by DFT. The HOS (Fig. 4d) and LUS (Fig. 4e) states exhibit enhanced spectral weight at the eight-membered ring sites and extended intensity bridging neighboring segments, indicating significant inter-dot hybridization. By contrast, the S1 (Fig. 4f) and C1 (Fig. 4g) excitations remain largely confined within individual segments, preserving spatial patterns consistent with standing-wave density modes and showing minimal evidence of coupling across the chain. This contrasting response demonstrates that inter-dot tunneling selectively reshapes electron-like states while leaving collective spin–charge excitations comparatively robust.

In a periodic array, coupling between neighboring units is expected to broaden discrete single-particle levels into miniband-like manifolds rather than producing simple bonding–antibonding splitting, consistent with the absence of clearly resolved doublets in the measured spectra. The observations therefore establish an interaction-dependent hierarchy of couplings: single-particle–like states (e.g., HOS and LUS) hybridize across atomically confined segments, whereas collective spin–charge separated excitations (e.g., C1 and S1) retain their identity even under atomic-scale connectivity, as illustrated in Fig. 5. This hierarchy reveals a fundamental distinction between single-particle and many-body responses under extreme quantum confinement.

**Discussion**

Collective quantum states are typically understood as emergent phenomena requiring spatial extensivity, with well-defined behavior arising only in spatially extended systems. By directly visualizing confined excitation modes in mirror twin boundary structures of monolayer $WSe_2$, our results demonstrate that this assumption is not necessary: the defining hallmark of Tomonaga–Luttinger liquid behavior, namely spin–charge separation, persists even when the effective system length approaches the nanometer scale (~ 1 nm). More importantly, we show that distinct classes of excitations exhibit qualitatively different responses to atomic confinement and coupling, revealing new emergent behavior beyond conventional single-particle intuition.

The atomically short mirror twin boundary segments investigated here represent a new class of many-body quantum dots, in which discrete excitation spectra arise from confined collective bosonic modes rather than single-particle electron states. Unlike conventional quantum dots governed primarily by single-electron quantization, these structures preserve interaction-driven dynamics characteristic of one-dimensional Tomonaga–Luttinger liquid behavior. The persistence of Luttinger-liquid behavior at near-atomic dimensions and the immunity to inter-dot coupling suggests that strong electronic correlations stabilize fractionalized collective excitations even under extreme spatial confinement.

The realization that atomically confined TLL quantum dots exhibit selective coupling behavior establishes a new platform for engineering correlated quantum matter from atomic-scale building blocks. In future, it would be intriguing to probe whether entanglement in a TLL reduces from the thermodynamic limit to the atomic limit and then becomes enriched in a chain of coupled many-body TLL quantum dots. In summary, our results demonstrate that collective excitations can remain well defined and rather robust even when spatial extent approaches atomic dimensions, thereby redefining expectations for the minimal size required to sustain many-body quantum behavior.

## Methods

### Sample growth for STM

High-quality buffer on SiC was synthesized using a two-step process. First, the monolayer epitaxial graphene was synthesized using silicon sublimation from the Si face of the semi-insulating SiC substrates (II–VI). Before the growth, the SiC substrates were annealed in 10% hydrogen (balance argon) at 1,500 °C for 30 min to remove subsurface damages due to chemical and mechanical polishing. Then monolayer epitaxial graphene (MLEG) was formed at 1,800 °C for 30 min in a pure argon atmosphere. Next, a Ni stressor layer was used to exfoliate the top graphene layer to obtain fresh and high-quality buffer on SiC. After this, 270 nm of Ni was e-beam deposited on MLEG at a rate of 5 Å $s^{-1}$ as a stressor layer. Then a thermal release tape was used to peel off the top graphene layer from the substrate. The growth of $WSe_2$ crystals on an epitaxial graphene substrate was carried out at 800 °C in a custom-built vertical cold-wall chemical vapor deposition (CVD) reactor for 20 min[49]. The tungsten hexacarbonyl ($W(CO)_6$) (99.99%, Sigma-Aldrich) source was kept inside a stainless-steel bubbler in which the temperature and pressure of the bubbler were always held at 37 °C and 730 torr respectively. Mass-flow controllers were used to supply $H_2$ carrier gas to the bubbler to transport the $W(CO)_6$ precursor into the CVD chamber. The flow rate of the $H_2$ gas through the bubbler was maintained at a constant 8 standard cubic centimeters per minute (sccm), which resulted in a $W(CO)_6$ flow rate of $9.0 \times 10^{-4}$ sccm at the outlet of the bubbler. $H_2Se$ (99.99%, Matheson) gas was supplied from a separate gas manifold and introduced at the inlet of the reactor at a constant flow rate of 30 sccm.

### DFT Calculations

All Density Functional Theory (DFT) calculations were calculated via the Quantum ESPRESSO suite[50,51]. The PBE exchange-correlation functional[52] was used in our calculations. The $WSe_2$ ribbons modeling twin-mirror boundary and four 4-ring boundary have length of 58 and 67 Å and contain 60 and 308 atoms, respectively. The structural optimization was obtained with criteria for force < 0.025 eV/°A, pressure < 0.5 kbar and total energy < 0.0014 eV. The density of states of four 4-ring boundary were calculated using k-mesh of 1 × 12 × 1. We employed optimized norm-conserving pseudopotentials[53] from the PseudoDojo library[54]. To reduce the computational cost, we used planewaves kinetic energy cutoff of 78 Ry which still guarantees fair precision[54]. The spin-orbit coupling effect was taken into account.

**STM and STS measurements**

STM and STS measurements on the MTBs in MOCVD grown $WSe_2$ were conducted in a home-built STM at 4.3 K in the ultra-high vacuum chamber, with a base pressure of $2.0\text{x}10^{-11}$ torr. The W tip was prepared by electrochemical etching and then cleaned by in-situ electron-beam heating. STM dI/dV spectra were measured using a standard lock-in technique, for which the modulation frequency was 758 Hz.

**Acknowledgement**

We would like to acknowledge the fruitful discussion with Charles Kane on the TLL couplings. F.Z., Y.L., H.K., YY. L. and C.-K.S. were primarily supported by the NSF through the Center for Dynamics and Control of Materials: an NSF Materials Research Science and Engineering Center under cooperative agreement no. DMR-2308817. F.Z., Y.L., H.K., YY. L. and C.-K.S. also acknowledge the support from the US Air Force grant no. FA2386-21-1-4061, NSF grant no. DMR-2219610, and the Welch Foundation F-2164. C. D. and J. A. R. were supported by the Penn State Center for Nanoscale Science (NSF grant no. DMR-2011839) and the Penn State 2DCC-MIP (NSF grant no. DMR-1539916). Y.-C.L. acknowledges the support from the Center for Emergent Functional Matter Science (CEFMS) of NYCU and the Yushan Young Scholar Program from the Ministry of Education of Taiwan. V.-A.H. and F.G. are suppoerted by the the Robert A. Welch Foundation under Award No. F-2139-20230405. Computational resources were provided by the National Energy Research Scientific Computing Center (a DOE Office of Science User Facility supported under Contract No. DE-AC02-05CH11231), the Argonne Leadership Computing Facility (a DOE Office of Science User Facility supported under Contract DE-AC02-06CH11357), and the Texas Advanced Computing Center (TACC) at The University of Texas at Austin. N.K.D. and F.Z. were supported by NSF under Grants No. DMR-2324033 and No. DMR-2414726 and by the Welch Foundation under Grant No. AT-2264-20250403.

## Author Contributions

F.Z. (UT Austin) and Y.L. carried out the STM and STS measurements under the supervision of C.-K.S. N.K.D performed the theoretical modeling under the supervision of F.Z. (UT Dallas) C.D. prepared the graphitic buffer layer/SiC and Y.-C.L. synthesized the $WSe_2$ flakes using MOCVD. J.A.R. supervised the sample synthesis. VA. H. performed the DFT calculations under the supervision of F.G. H.K. and YY. L. refined sample cleaning in UHV. F.Z. (UT Austin), Y.L., and C.-K.S. analyzed the STM data. F.Z. (UT Austin), and C.-K.S. wrote the paper with contributions from all the authors. † These authors contributed equally to this work.

**Competing Interests:** All the authors declare no competing interests.

## Data availability

Source data that reproduce the plots in the main text are provided with this paper. Source data that reproduce the plots in the Supplementary Information are available on request.

## References

1. Anderson, P. W. More Is Different: Broken symmetry and the nature of the hierarchical structure of science. *Science* **177**, 393–396 (1972).
2. Von Delft, J. & Ralph, D. C. Spectroscopy of discrete energy levels in ultrasmall metallic grains. *Physics Reports* **345**, 61–173 (2001).
3. Lee, P. A. & Ramakrishnan, T. V. Disordered electronic systems. *Rev. Mod. Phys.* **57**, 287–337 (1985).
4. Brus, L. E. Electron–electron and electron-hole interactions in small semiconductor crystallites: The size dependence of the lowest excited electronic state. *The Journal of Chemical Physics* **80**, 4403–4409 (1984).
5. Efros, Al. L. & Rosen, M. The Electronic Structure of Semiconductor Nanocrystals. *Annu. Rev. Mater. Sci.* **30**, 475–521 (2000).

6. Eggert, S., Johannesson, H. & Mattsson, A. Boundary Effects on Spectral Properties of Interacting Electrons in One Dimension. *Phys. Rev. Lett.* **76**, 1505–1508 (1996).
7. Barak, G. *et al.* Interacting electrons in one dimension beyond the Luttinger-liquid limit. *Nature Phys* **6**, 489–493 (2010).
8. Anfuso, F. & Eggert, S. Luttinger liquid in a finite one-dimensional wire with box-like boundary conditions. *Phys. Rev. B* **68**, 241301 (2003).
9. Kakashvili, P., Johannesson, H. & Eggert, S. Local spectral weight of a Luttinger liquid: Effects from edges and impurities. *Phys. Rev. B* **74**, 085114 (2006).
10. Mattsson, A. E., Eggert, S. & Johannesson, H. Properties of a Luttinger liquid with boundaries at finite temperature and size. *Phys. Rev. B* **56**, 15615–15628 (1997).
11. Giamarchi, T. *Quantum Physics in One Dimension*. (Oxford University Press, 2003). doi:10.1093/acprof:oso/9780198525004.001.0001.
12. J Voit. One-dimensional Fermi liquids. *Reports on Progress in Physics* **58**, 977 (1995).
13. Tomonaga, S. Remarks on Bloch's Method of Sound Waves applied to Many-Fermion Problems. *Progress of Theoretical Physics* **5**, 544–569 (1950).
14. Luttinger, J. M. An Exactly Soluble Model of a Many-Fermion System. *Journal of Mathematical Physics* **4**, 1154–1162 (1963).
15. F D M Haldane. 'Luttinger liquid theory' of one-dimensional quantum fluids. I. Properties of the Luttinger model and their extension to the general 1D interacting spinless Fermi gas. *Journal of Physics C: Solid State Physics* **14**, 2585 (1981).
16. Fiete, G. A. *Colloquium* : The spin-incoherent Luttinger liquid. *Rev. Mod. Phys.* **79**, 801–820 (2007).

17. Kane, C., Balents, L. & Fisher, M. P. A. Coulomb Interactions and Mesoscopic Effects in Carbon Nanotubes. *Phys. Rev. Lett.* **79**, 5086–5089 (1997).

18. Bockrath, M. *et al.* Luttinger-liquid behaviour in carbon nanotubes. *Nature* **397**, 598–601 (1999).

19. Ishii, H. *et al.* Direct observation of Tomonaga–Luttinger-liquid state in carbon nanotubes at low temperatures. *Nature* **426**, 540–544 (2003).

20. Postma, H. W. Ch., Teepen, T., Yao, Z., Grifoni, M. & Dekker, C. Carbon Nanotube Single-Electron Transistors at Room Temperature. *Science* **293**, 76–79 (2001).

21. Wang, S. *et al.* Nonlinear Luttinger liquid plasmons in semiconducting single-walled carbon nanotubes. *Nat. Mater.* **19**, 986–991 (2020).

22. Auslaender, O. M. *et al.* Spin-Charge Separation and Localization in One Dimension. *Science* **308**, 88–92 (2005).

23. Kim, B. J. *et al.* Distinct spinon and holon dispersions in photoemission spectral functions from one-dimensional SrCuO2. *Nature Phys* **2**, 397–401 (2006).

24. Jompol, Y. *et al.* Probing Spin-Charge Separation in a Tomonaga-Luttinger Liquid. *Science* **325**, 597–601 (2009).

25. Jia, J. *et al.* Tuning the many-body interactions in a helical Luttinger liquid. *Nat Commun* **13**, 6046 (2022).

26. Bouchoule, I. *et al.* Platforms for the realization and characterization of Tomonaga–Luttinger liquids. *Nat Rev Phys* **7**, 565–580 (2025).

27. Batzill, M. Mirror twin grain boundaries in molybdenum dichalcogenides. *J. Phys.: Condens. Matter* **30**, 493001 (2018).

28. Jolie, W. *et al.* Tomonaga-Luttinger Liquid in a Box: Electrons Confined within MoS 2 Mirror-Twin Boundaries. *Phys. Rev. X* **9**, 011055 (2019).

29. Rossi, A. *et al.* Graphene-driven correlated electronic states in one dimensional defects within WS2. *Nat Commun* **16**, 5809 (2025).

30. Xia, Y. *et al.* Charge Density Modulation and the Luttinger Liquid State in $MoSe_2$ Mirror Twin Boundaries. *ACS Nano* **14**, 10716–10722 (2020).

31. Zhu, T. *et al.* Imaging gate-tunable Tomonaga–Luttinger liquids in 1H-MoSe2 mirror twin boundaries. *Nat. Mater.* **21**, 748–753 (2022).

32. Bagchi, M. *et al.* Spin-Polarized Scanning Tunneling Microscopy Measurements of an Anderson Impurity. *Phys. Rev. Lett.* **133**, 246701 (2024).

33. Deng, B. *et al.* Epitaxially Defined Luttinger Liquids on MoS 2 Bicrystals. *Phys. Rev. Lett.* **134**, 046301 (2025).

34. Murray, C. *et al.* Band Bending and Valence Band Quantization at Line Defects in MoS2. *ACS Nano* **14**, 9176–9187 (2020).

35. Komsa, H. & Krasheninnikov, A. V. Engineering the Electronic Properties of Two-Dimensional Transition Metal Dichalcogenides by Introducing Mirror Twin Boundaries. *Adv Elect Materials* **3**, 1600468 (2017).

36. Pathirage, V., Lasek, K., Krasheninnikov, A. V., Komsa, H. P. & Batzill, M. Mirror twin boundaries in WSe2 induced by vanadium doping. *Materials Today Nano* **22**, 100314 (2023).

37. Van Efferen, C. *et al.* Modulated Kondo screening along magnetic mirror twin boundaries in monolayer MoS2. *Nat. Phys.* **20**, 82–87 (2024).

38. Ahn, H. *et al.* Integrated 1D epitaxial mirror twin boundaries for ultrascaled 2D MoS2 field-effect transistors. *Nat. Nanotechnol.* **19**, 955–961 (2024).

39. Yang, X. *et al.* Manipulating Hubbard-type Coulomb blockade effect of metallic wires embedded in an insulator. *National Science Review* **10**, nwac210 (2023).

40. Pan, Z. *et al.* Ferromagnetism and correlated insulating states in monolayer Mo33Te56. *Nature Communications* **16**, 3084 (2025).

41. Mukhopadhyay, R., Kane, C. L. & Lubensky, T. C. Crossed sliding Luttinger liquid phase. *Phys. Rev. B* **63**, 081103 (2001).

42. Furukawa, S. & Kim, Y. B. Entanglement entropy between two coupled Tomonaga-Luttinger liquids. *Phys. Rev. B* **83**, 085112 (2011).

43. Yoshioka, H. & Suzumura, Y. Properties of fluctuations in two coupled chains of luttinger liquids. *J Low Temp Phys* **106**, 49–72 (1997).

44. Fendley, P. & Nayak, C. Tunneling between Luttinger liquids. *Phys. Rev. B* **63**, 115102 (2001).

45. Vishwanath, A. & Carpentier, D. Two-Dimensional Anisotropic Non-Fermi-Liquid Phase of Coupled Luttinger Liquids. *Phys. Rev. Lett.* **86**, 676–679 (2001).

46. Fiete, G. A., Le Hur, K. & Balents, L. Coulomb drag between two spin-incoherent Luttinger liquids. *Phys. Rev. B* **73**, 165104 (2006).

47. Wang, P. *et al.* One-dimensional Luttinger liquids in a two-dimensional moiré lattice. *Nature* **605**, 57–62 (2022).

48. Li, H. *et al.* Imaging tunable Luttinger liquid systems in van der Waals heterostructures. *Nature* **631**, 765–770 (2024).

49. Lin, Y.-C. *et al.* Realizing Large-Scale, Electronic-Grade Two-Dimensional Semiconductors. *ACS Nano* **12**, 965–975 (2018).

50. Giannozzi, P. *et al.* QUANTUM ESPRESSO: a modular and open-source software project for quantum simulations of materials. *J. Phys.: Condens. Matter* **21**, 395502 (2009).

51. Giannozzi, P. *et al.* Advanced capabilities for materials modelling with Quantum ESPRESSO. *J. Phys.: Condens. Matter* **29**, 465901 (2017).

52. Perdew, J. P., Burke, K. & Ernzerhof, M. Generalized Gradient Approximation Made Simple. *Phys. Rev. Lett.* **77**, 3865–3868 (1996).

53. Hamann, D. R. Optimized norm-conserving Vanderbilt pseudopotentials. *Phys. Rev. B* **88**, 085117 (2013).

54. Van Setten, M. J. *et al.* The PseudoDojo: Training and grading a 85 element optimized norm-conserving pseudopotential table. *Computer Physics Communications* **226**, 39–54 (2018).

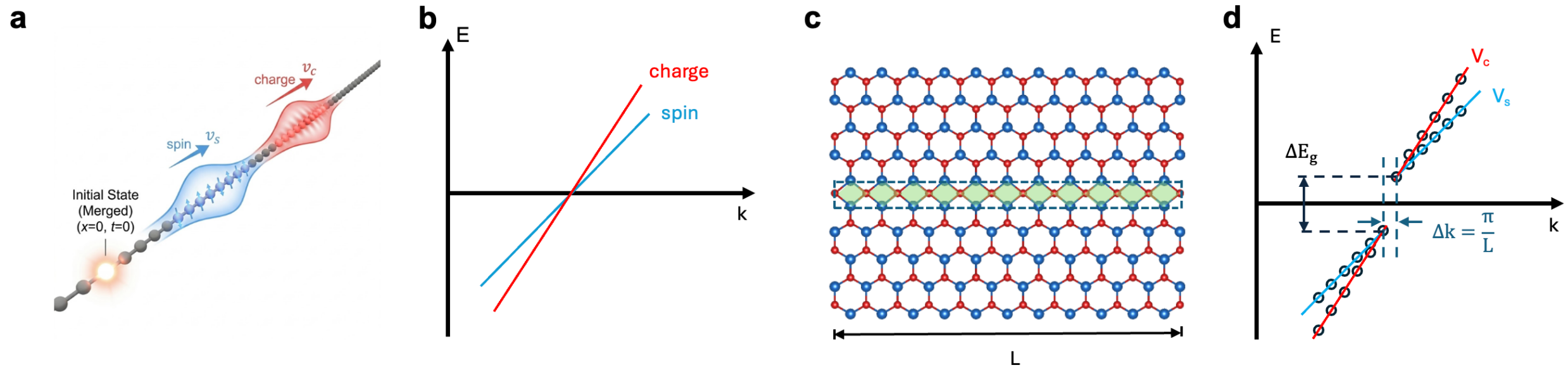

**Figure 1 | Spin-Charge separation in 1D Tomonaga-Luttinger liquid.** (a) Schematic of the real-space spin and charge excitations in 1D TLL. (b) Schematic of spin and charge excitations in an infinite long 1D TLL system. (c) Schematic of an MTB with a finite length L in monolayer $WSe_2$. (d) Schematic Luttinger-liquid-in-a-box spectrum showing the spin and charge excitations in a finite-sized 1D TLL system.

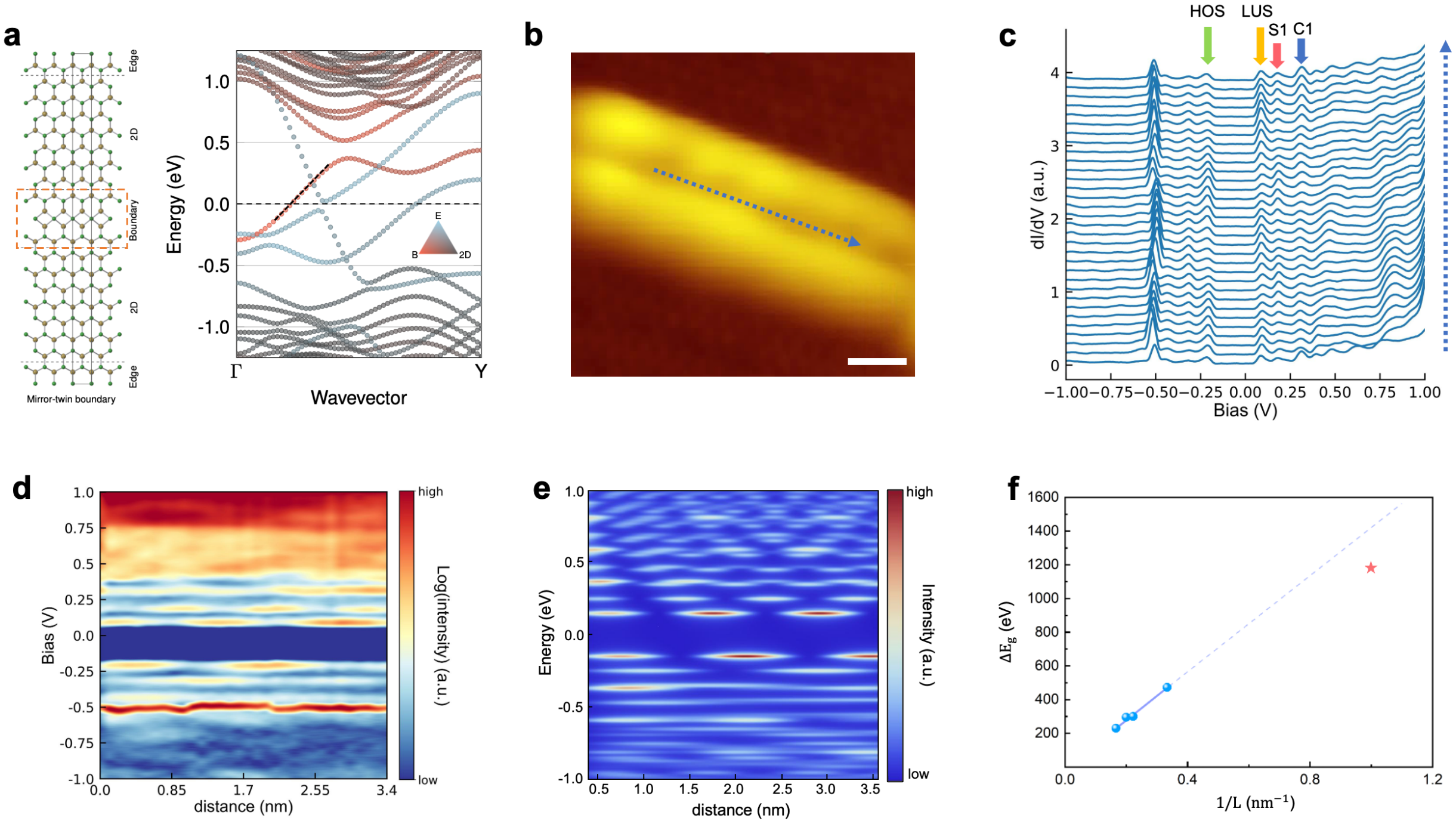

**Figure 2 | Tomonaga-Luttinger liquid behavior in $WSe_2$ MTB.** (a) Left: schematic model of MTB. W atoms, brown; Se atoms, green. Right: Calculated band structure for the ribbon geometry of (a) with a periodic boundary condition in the direction along the MTB. Horizontal dashed line denotes the position of $E_F$ at $E = 0$. The band present at the MTB is colored orange and crosses $E_F$ at $k_F \approx \pi/4a$. Monolayer 2D bands are colored grey while bands located at the ribbon edges specific to the finite-sized supercell are colored light blue. For labeling: B stand for boundary, E stand for edge. (b) STM topography image of a 5-nm long MTB. Scale bar: 1 nm. (c) STS acquired along the MTB with the path marked by the blue arrow shown in (b). (d) Corresponding dI/dV signal as a function of energy and position (Vs = 1 V, I = 50 pA). (e) Theoretical LDOS predicted by the finite TLL model using $\Delta E_g = 0.3\ eV$, $K_c = 0.46$, $K_s = 1$, $K_F^+ = 6\pi/L$, $K_F^- = 5\pi/L$, $\frac{v_c\pi}{L} = 0.22\ eV$ and $\frac{v_s\pi}{L} = 0.10\ eV$. $E_{gap}$ is the energy gap at the Fermi level. $K_c/K_s$ is the charge/spin-channel Luttinger parameter. $v_s$ and $v_c$ are respectively velocities for the spin and charge excitations. (f) Fitting of MTB with different lengths, showing a linear dependence of the bandgap on the inverse MTB length.

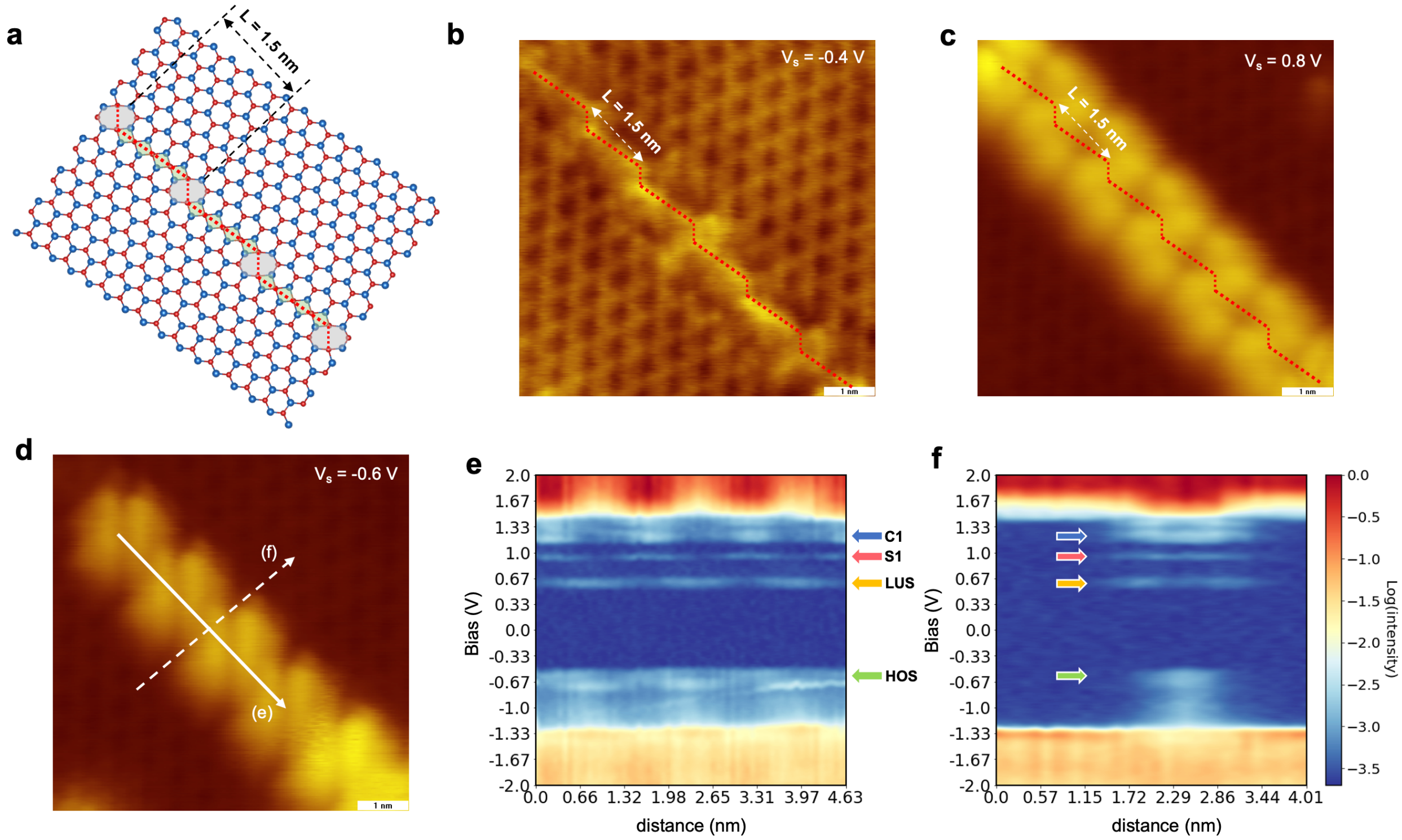

**Figure 3 | Structure for a chain of Tomonaga-Luttinger liquid quantum dots.** (a) Schematic model for a periodic chain of TLL quantum dots. The red dashed lines mark individual one-nanometer-long TLL quantum dots connected by the 8-membered ring structures. The period marked by the black dashed arrow is $\sqrt{21}a$ = 1.5 nm. (b) Inside the energy gap $\Delta E_g$, an atomically resolved STM image showing the structure of a periodic chain of TLL quantum dots (Vs = -0.4 V, I = -20 pA). (c,d) STM images revealing the chain periodicity of 1.5 nm at (c) (Vs = 0.8 V, I = 20 pA) and (d) (Vs = -0.6 V, I = -20 pA). (e,f) Color maps of the dI/dV spectra (e) along and (f) across the TLL chain.

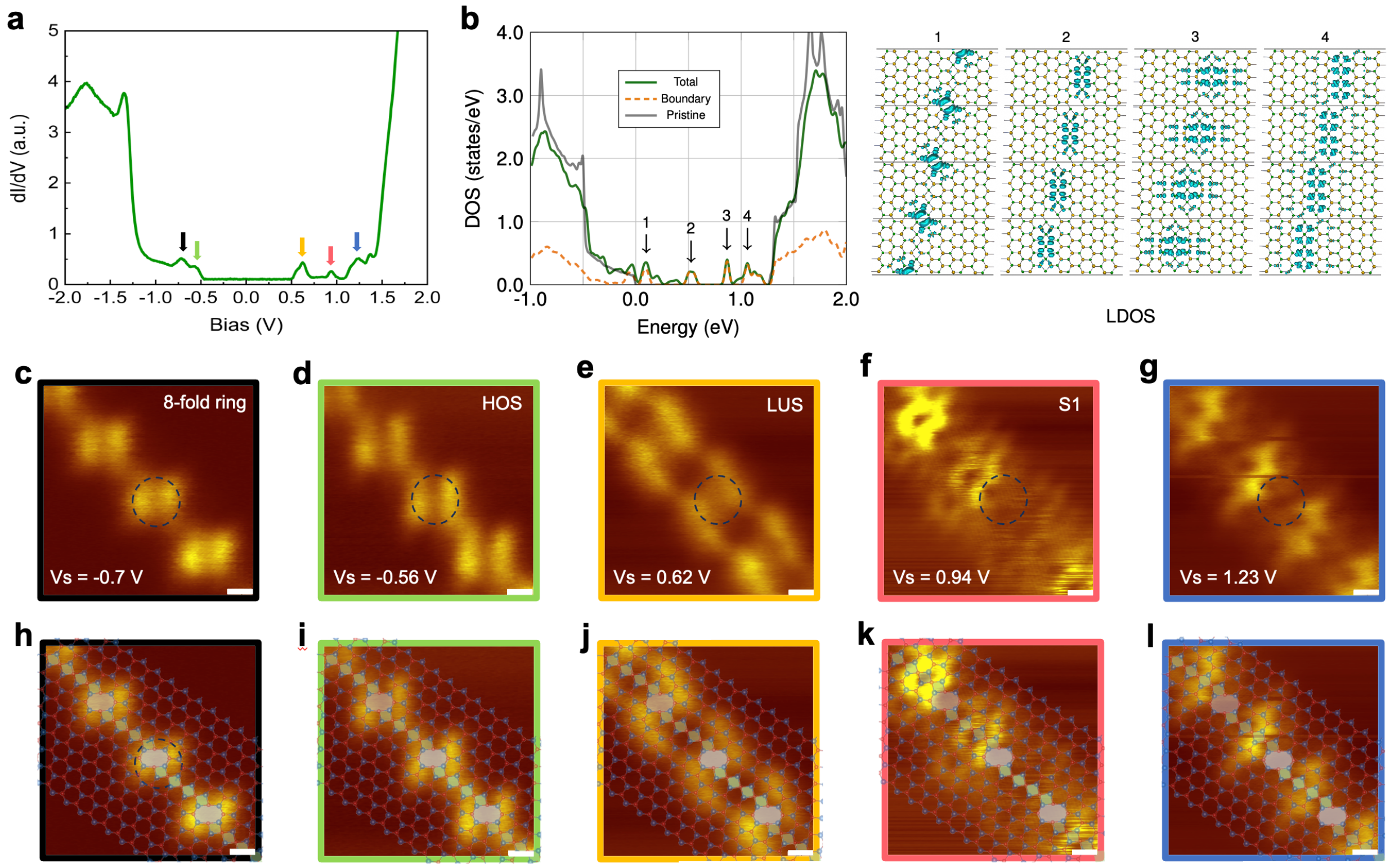

**Figure 4 | Inter-QD and intra-QD couplings in the TLL QD chain.** (a) Averaged result of a line STS along the TLL QD chain. (b) Green curve shows the calculated density of states of the TLL QD chain, while the black curve corresponds to the DOS of pristine monolayer $WSe_2$. The spatial distribution of DOS at energy peaks 1-4 are respectively shown in the right panels. (c-g) Constant height dI/dV mapping of the chain of TLL QDs at sample bias (c) Vs = -0.7 V; (d) Vs = -0.56 V; (e) Vs = 0.62 V; (f) Vs = 0.94 V; (g) Vs = 1.23 V. The corresponding modes at these energy levels are marked by the colored arrows in (a). (h-l) The same constant-height dI/dV maps as in (c–g), overlaid with a schematic illustration of the TLL QD chain. The eight-membered ring structure is marked with a black dash circle. Scale bars: 500 nm.

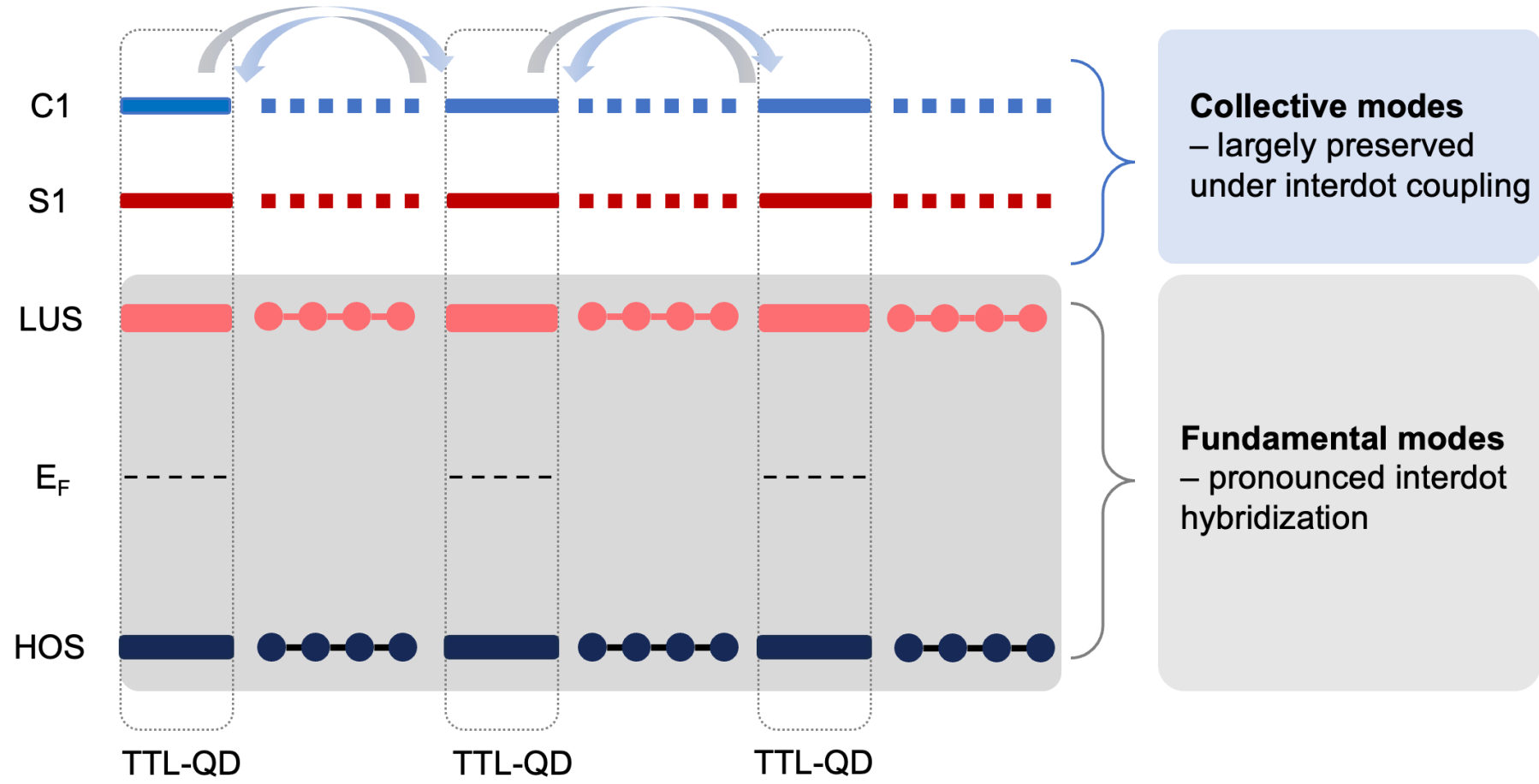

**Figure 5 | Interaction-dependent hierarchy of couplings in the atomic-limit TLL quantum-dot chain.** Fundamental electron-like modes (HOS/LUS) exhibit pronounced inter-dot hybridization across neighboring segments, whereas collective spin–charge excitations (S1/C1) remain largely confined within individual segments, preserving their standing-wave character. The schematic summarizes the distinct coupling responses revealed by spectroscopic imaging.